\documentclass{article}

\usepackage{microtype}
\usepackage{graphicx}
\usepackage{subfigure}
\usepackage{booktabs} 

\usepackage{hyperref}

\usepackage[accepted]{icml2024}

\usepackage{amsmath}
\usepackage{amssymb}
\usepackage{mathtools}
\usepackage{amsthm}
\usepackage[T1]{fontenc}

\usepackage[capitalize,noabbrev]{cleveref}

\theoremstyle{plain}

\theoremstyle{definition}

\theoremstyle{remark}

\usepackage[textsize=tiny]{todonotes}

\icmltitlerunning{Matching experts by training from scratch}

\begin{document}

\twocolumn[
\icmltitle{Matching domain experts by training from scratch on domain knowledge}



\icmlsetsymbol{equal}{*}

\begin{icmlauthorlist}
\icmlauthor{Xiaoliang Luo}{ucl}
\icmlauthor{Guangzhi Sun}{cam}
\icmlauthor{Bradley C. Love}{ucl,turing}
\end{icmlauthorlist}

\icmlaffiliation{ucl}{Department of Experimental Psychology, University College London, UK}
\icmlaffiliation{cam}{Department of Engineering, University of Cambridge, UK}
\icmlaffiliation{turing}{The Alan Turing Institute, UK}

\icmlcorrespondingauthor{Xiaoliang Luo}{xiao.luo.17@ucl.ac.uk}

\icmlkeywords{Large Language Models, Neuroscience, Scientific Discovery}

\vskip 0.3in
]



\printAffiliationsAndNotice{}  

\begin{abstract}
Recently, large language models (LLMs) have outperformed human experts in predicting the results of neuroscience experiments \citep{luo_large_2024}. What is the basis for this performance? One possibility is that statistical patterns in that specific scientific literature, as opposed to emergent reasoning abilities arising from broader training, underlie LLMs' performance.
To evaluate this possibility, we trained (next word prediction) a relatively small 124M-parameter GPT-2 model on 1.3 billion tokens of domain-specific knowledge. Despite being orders of magnitude smaller than larger LLMs trained on trillions of tokens, small models achieved expert-level performance in predicting neuroscience results. Small models trained on the neuroscience literature succeeded when they were trained from scratch using a tokenizer specifically trained on neuroscience text or when the neuroscience literature was used to finetune a pretrained GPT-2. Our results indicate that expert-level performance may be attained by even small LLMs through domain-specific, auto-regressive training approaches.

\end{abstract}

\section{Introduction}
Large language models (LLMs) are statistical machines typically designed to predict the next token—whether it's a word, pixel, or protein sequence. Leveraging vast amounts of training data, LLMs have demonstrated impressive capabilities, including passing professional exams, reasoning (though with limitations), translation, solving mathematics problems, and writing computer code \cite{strack_visual_2023, srivastava_beyond_2022, gunasekar_textbooks_2023}.

Traditionally, the human-level performance of large language models (LLMs) has been evaluated using benchmarks that focus on their \textit{backward-looking} capabilities, such as core knowledge retrieval and reasoning within a given context. Notable benchmarks include MMLU \cite{hendrycks_measuring_2021}, PubMedQA \cite{jin_pubmedqa_2019}, and MedMCQA \cite{pal_medmcqa_2022}. However, recent research by \citet{luo_large_2024} has highlighted LLMs' exceptional forward-looking capabilities, particularly in predicting novel outcomes of neuroscience studies. With the development of BrainBench, a \textit{forward-looking} neuroscience benchmark, \citet{luo_large_2024} have shown that LLMs can outperform neuroscientists in predicting the results of neuroscientific experiments when provided with the experiment's background and methodologies. These findings raise important questions about the nature of scientific progress, suggesting that many discoveries might largely be iterations of noisy signals from decades of scientific literature. Additionally, they prompt a reevaluation of the extent to which accurate predictions of the future rely more on pattern recognition by auto-regressive models than on traditional scientific reasoning.

In this contribution, we explore the effects of training on domain-specific data by employing a significantly smaller language model, GPT-2 with 124 million parameters \citep{radford_language_2019}, on a neuroscience-focused dataset containing 1.3 billion tokens. This approach helps assess the effectiveness of auto-regressive training on specialized data in approximating human-level performance. Despite the model size being only about $0.056\%$ to $1\%$\footnote{Estimated using 7B and 180B LLMs.} of those evaluated by \citet{luo_large_2024} and the training data being about $0.065\%$\footnote{Estimated based on reported Llama-2 training data size (2 trillion tokens).} of those used in \citet{luo_large_2024}, we show both finetuning a pretrained 124M-parameter GPT-2 and training it from scratch with a custom tokenizer for neuroscience yield models that achieve $63.5\%$ and $63\%$ accuracy on BrainBench, matching the performance of human experts ($63.4\%$). Larger GPT-2 variant (774M) pretrained on the same data yield even stronger BrainBench performance (surpassing human experts; Fig \ref{fig:model_vs_human}) with improved confidence calibration beneficial for human-model teaming (see Appendix \ref{app:calibration}, \ref{app:overall}).

\section{Method}
\subsection{BrainBench}
BrainBench has curated 200 test cases from abstracts in the \textit{Journal of Neuroscience published in 2023}. These abstracts are categorized into five sections: Behavioral/Cognitive, Systems/Circuits, Neurobiology of Disease, Development/Plasticity/Repair, and Cellular/Molecular.

Each test case includes a published abstract alongside a modified version crafted by neuroscientists. These modifications, though minimal, significantly alter the results—for instance, by changing the roles of brain regions or reversing a result's direction (e.g., from "decreases" to "increases"). Despite these changes, the altered abstracts remain logically coherent.

The test-taker's challenge is to identify the correct study outcome by choosing between the original abstract and its altered counterpart.

\subsection{Model evaluation}
We presented models with two versions of the abstracts from each test case separately. We prefixed each abstract with the prompt ``\textit{You are a neuroscientist with deep knowledge in neuroscience. Here is an abstract from a neuroscience publication:}''. We then measured the perplexity of both passages and used perplexity as the indicator of whether models favor one abstract or the other.

Perplexity measures the degree of uncertainty of a model when generating a particular sequence of text and is defined as the exponentiated average negative log-likelihood of a tokenized sequence. If we have a tokenized abstract \( X = (x_0, x_1, \ldots, x_t) \), then the perplexity of \( X \), given a model parameterized by $\theta$ is,

\begin{equation}
    PPL(X) = \exp \left\{ -\frac{1}{t} \sum_{i}^{t} \log p_\theta (x_i | x_{<i}) \right\}
\end{equation}

where \( \log p_\theta (x_i | x_{<i}) \) is the log-likelihood of the \( i \)th token conditioned on the preceding tokens \( x_{<i} \) according to the model. Given both the original and the altered abstracts, we used the abstract with lower perplexity as the model's decision and evaluated the overall accuracy across the entire BrainBench dataset accordingly.

\subsection{Human evaluation}
Previous work \cite{luo_large_2024} collected human judgements from 171 neuroscience experts on BrainBench. These data are publicly available\footnote{https://github.com/braingpt-lovelab/BrainBench} and provide a useful comparison to LLM performance.

\subsection{Model configurations}
We considered a number of variants of GPT-2 differ by their training strategies including training data and tokenization. Model variants are summarized in Table \ref{tab:varaints}.

\begin{table}[t]
\vskip -0.15in
\caption{\textbf{Model variants.}}
\label{tab:varaints}
\vskip -0.15in
\begin{center}
\begin{small}
\begin{tabular}{lcccr}
\toprule
Variant & Training & Data & Tokenizer \\
\midrule
GPT2-Untrained    & - & - & pretrained \\
GPT2-Pretrained & from scratch & WebText & pretrained\\
GPT2-Scratch    & from scratch & neuroscience & pretrained \\
GPT2-Finetuned & finetune & neuroscience & pretrained\\
GPT2-Neuro & from scratch & neuroscience & custom \\
\bottomrule
\end{tabular}
\end{small}
\end{center}
\vskip -0.3in
\end{table}

The pretrained GPT-2 and the tokenizer were loaded from Huggingface hub\footnote{https://huggingface.co/openai-community/gpt2}, which were trained on the WebText dataset collected by OpenAI \cite{radford_language_2019}. The neuroscience training data was collected by \citet{luo_large_2024} (see Sec. \ref{training_data}). The models trained from scratch and finetuned used the neuroscience data only. The neuro-tokenizer employs GPT-2's tokenization strategy \citep{radford_language_2019}, adapted from Byte Pair Encoding (BPE) \citep{gage_new_1994} for word segmentation \citep{sennrich_neural_2016}. It's trained anew on neuroscience data used for model training, maintaining a vocabulary size of 50,257 tokens. 

\subsection{Neuroscience training data}
\label{training_data}
The data we used to train GPT-2 from scratch, finetune the pretrained GPT-2 as well as train the neuro-tokenizer were collected by \citet{luo_large_2024}. The training data spans Neuroscience publication (abstracts and full articles) dates 2002-2022, totaling
1.3 billion tokens. We randomly allocated 90\% of the data for training, reserving the remaining 10\% for validation. Training details see Appendix.

\section{Results}
\begin{figure}
\begin{center}
\centerline{\includegraphics[width=.8\columnwidth]{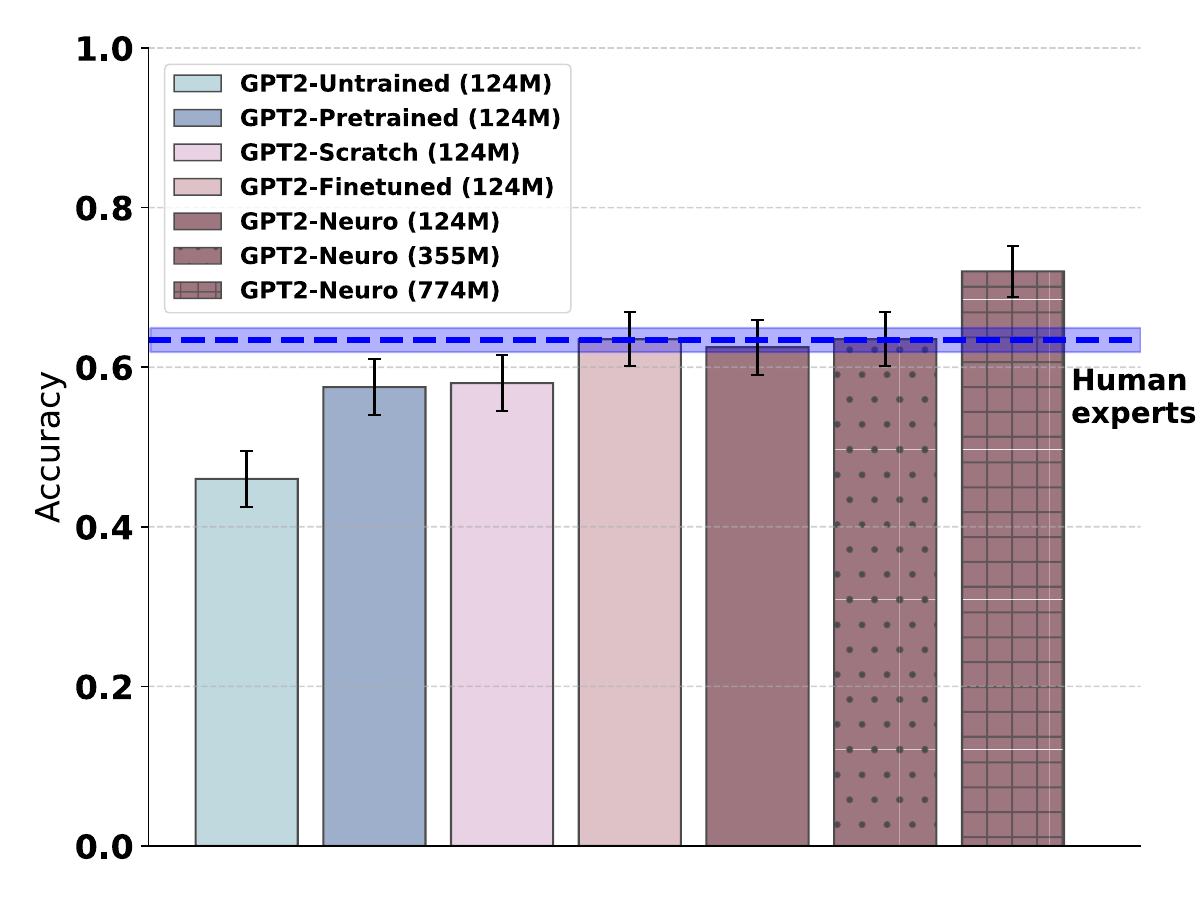}}
\caption{\textbf{Performance of human experts and models on BrainBench.} 
Two configurations of GPT-2 models achieve human-level performance on BrainBench: one by fine-tuning the pretrained GPT-2 on neuroscience literature, and the other by using a new tokenizer (GPT2-Neuro) trained on neuroscience literature and retraining GPT-2 from scratch with only neuroscience data. Versions of GPT-2 that are untrained, pretrained, or trained solely on neuroscience data without these modifications underperform compared to experts on BrainBench. Larger GPT-2 (774M) trained using neuroscience literature surpassed human-level performance. }
\label{fig:model_vs_human}
\end{center}
\vskip -.3in
\end{figure}

We explored various training strategies and found that finetuning the pretrained GPT-2 on 20 years of neuroscience literature allowed it to achieve human-level performance on BrainBench, recording a 63.5\% accuracy (Fig. \ref{fig:model_vs_human}; human experts: 63.4\%). Training GPT-2 from scratch solely with neuroscience literature was less effective. However, developing a new tokenizer tailored to neuroscience literature and using it to retrain GPT-2 from scratch with the same data resulted in a performance on par with human experts, achieving 63\% accuracy (Fig. \ref{fig:model_vs_human}). Notably, the amount of domain-specific data used to train GPT-2 from scratch is only about one-seventh of the text used to pretrain the original model. This indicates two effective approaches to reach human-level performance: pretraining on a broad general corpus followed by finetuning on domain-specific data, or using a specialized tokenizer and significantly less domain-specific data.

\begin{figure}
\begin{center}
\centerline{\includegraphics[width=.8\columnwidth]{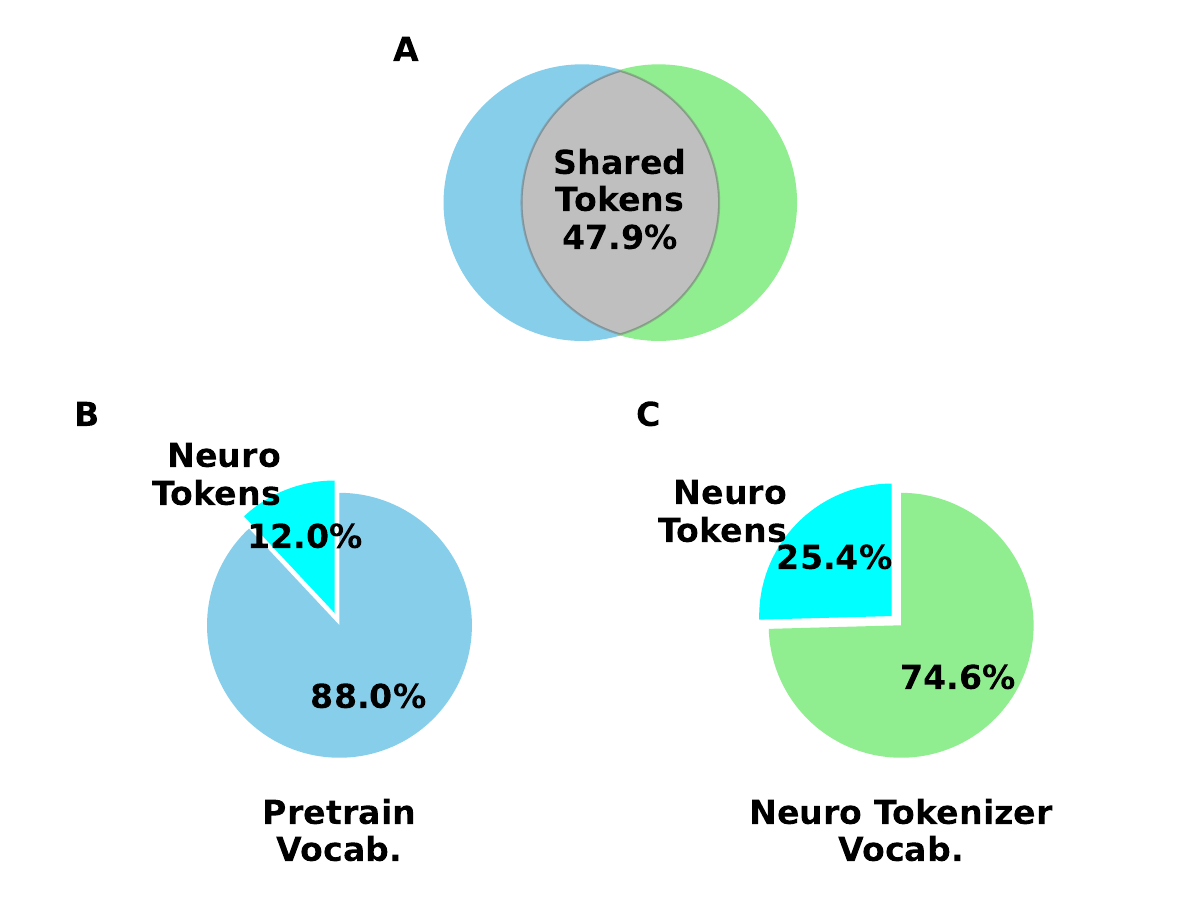}}
\caption{\textbf{Token Analysis.} 
(A) The pretrained GPT-2 tokenizer and the neuro-tokenizer share 47.9\% of their vocabularies. (B-C) Of the vocabularies from the two tokenizers, 12.0\% of the tokens from the pretrained tokenizer are commonly associated with neuroscience according to GPT-4, compared to 25.4\% of the tokens from the neuro-tokenizer.}
\label{fig:tokens}
\end{center}
\vskip -.3in
\end{figure}

To assess the impact of a specialized tokenizer, we compared the tokens generated by the pretrained GPT-2 tokenizer with those from our neuro-tokenizer, trained on neuroscience data. The two tokenizers shared 47.9\% of their vocabularies (Fig \ref{fig:tokens}A). We utilized GPT-4 (zero-shot prompting) to analyze each vocabulary and identify tokens frequently associated with neuroscience. Our findings showed that the neuro-tokenizer contained twice the proportion of neuroscience-related tokens compared to the pretrained tokenizer (Fig. \ref{fig:tokens}B-C). This significant improvement in specialized tokenization suggests that it is possible to pretrain GPT-2 from scratch with significantly less neuroscience data using the neuro-tokenizer, yet achieve performance comparable to both the finetuned model and human experts.

To better understand the differences in tokenization by the two tokenizers, we analyzed examples from BrainBench test cases where the pretrained GPT-2 answered incorrectly, whereas the GPT-2 trained with the neuro-tokenizer responded correctly. Figure \ref{fig:tokens_viz} illustrates how the neuro-tokenizer more effectively preserves domain-specific terminologies, such as brain regions or neurotransmitters. We believe that this specialized tokenization allows the model to utilize limited domain knowledge more effectively and to consider a broader context within the fixed context window of the training data.

\section{Discussion}
In this contribution, we demonstrated that training a relatively small LLM (GPT-2) on limited domain-specific data can match the predictive performance of human experts on BrainBench. By finetuning GPT-2 with just a fraction of its pretraining data, we elevated its performance to the level of trained neuroscientists. Additionally, we showed that training GPT-2 from scratch, with domain-specific knowledge incorporated into the tokenizer, yields comparable results. This highlights the importance of preserving domain-specific terminologies during tokenization to improve language models' performance on specialized tasks, as suggested by \citet{yang_exkidneybert_2024} in the clinical science domain. Pretraining on small-scale knowledge with specialized tokenization offers a more efficient method for achieving human-like performance on domain-specific tasks. In addition, the resulting models show good calibration and ability to integrate information across context (Appendix \ref{app:calibration}, \ref{app:context}).

Achieving parity with and surpassing human experts using our simplified setup prompts questions about the essence of scientific progress. It suggests that a statistical machine, even one as basic as predicting the next word, can discern the intricate structure of a knowledge-rich field. Despite a significant performance gap (15\%) between GPT-2 (124M) and more advanced LLMs on BrainBench tests, larger GPT-2 models—still much smaller than their counterparts—are narrowing this gap when pretrained with neuroscience data (Appendix \ref{app:overall}). We suspect that the performance is influenced by both model size and the quality and relevance of the training data (see \citealt{luo_large_2024} for a discussion).

Working with smaller models does have benefits, such as enabling teams with modest resources to have full control over the training procedure. This control can minimize the risk of leakage and allow for additional hypotheses to be evaluated. For instance, in future work, we will evaluate whether training on adjacency fields like psychology impacts performance on BrainBench, a neuroscience benchmark. That degree of control is not possible using pretrained LLMs and will allow us to evaluate the structure of scientific disciplines.

\section*{Software and Data}
Human participant data, and intermediate data generated via simulations and analyses are publicly available at \href{https://github.com/braingpt-lovelab/BrainBench}{https://github.com/braingpt-lovelab/BrainBench}. Model weights and training data are available at \href{https://huggingface.co/BrainGPT}{https://huggingface.co/BrainGPT}.

All computer code associated with this work including model training, evaluation, data processing and analyses are publicly available at \href{https://github.com/braingpt-lovelab/matching_experts}{https://github.com/braingpt-lovelab/matching\_experts}.


\section*{Impact Statement}
This paper presents work whose goal is to advance the field of 
Machine Learning. There are many potential societal consequences 
of our work, none which we feel must be specifically highlighted here.

\section*{Acknowledgements}
This work was supported the ESRC (ES/W007347/1), Microsoft (Accelerate Foundation Models Research Program), and a Royal Society Wolfson Fellowship (18302) to B.C.L. 

\bibliography{references-ken,references-brad}
\bibliographystyle{icml2024}
\newpage
\appendix
\onecolumn

\section{Comparison between pretrained and neuro-tokenizer}

\begin{figure}[H]
\begin{center}
\centerline{\includegraphics[scale=0.5]{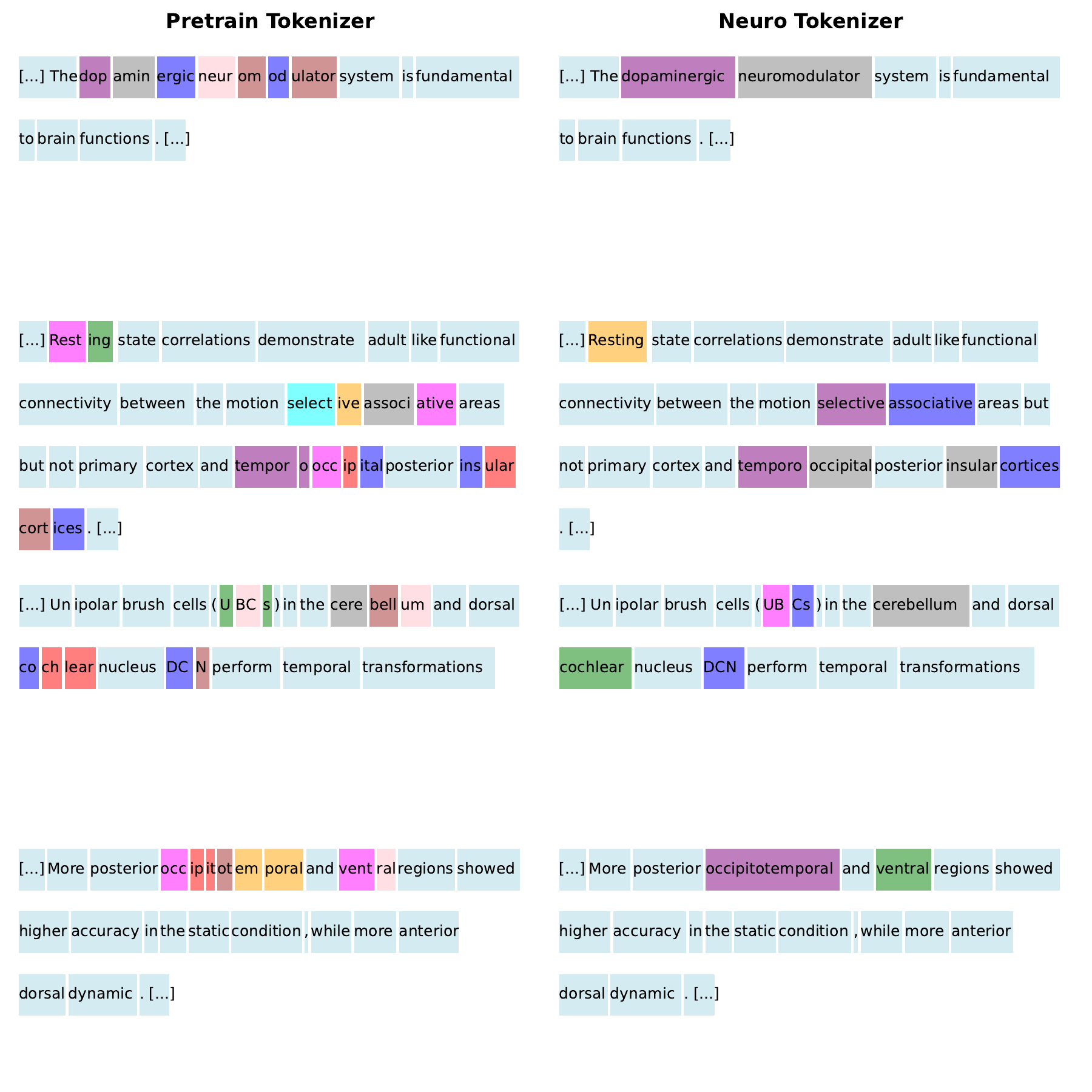}}
\caption{\textbf{Tokenization examples.} Compared to a pretrained tokenizer trained on general text, a neuro-tokenizer trained specifically on neuroscience literature better preserves domain-specific terminology, such as brain regions, in neuroscience.}
\label{fig:tokens_viz}
\end{center}
\vskip -0.2in
\end{figure}

\newpage
\section{Confidence calibration of domain-specific pretrained models}
\label{app:calibration}
\begin{figure}[H]
    \centering
    \includegraphics[width=\textwidth]{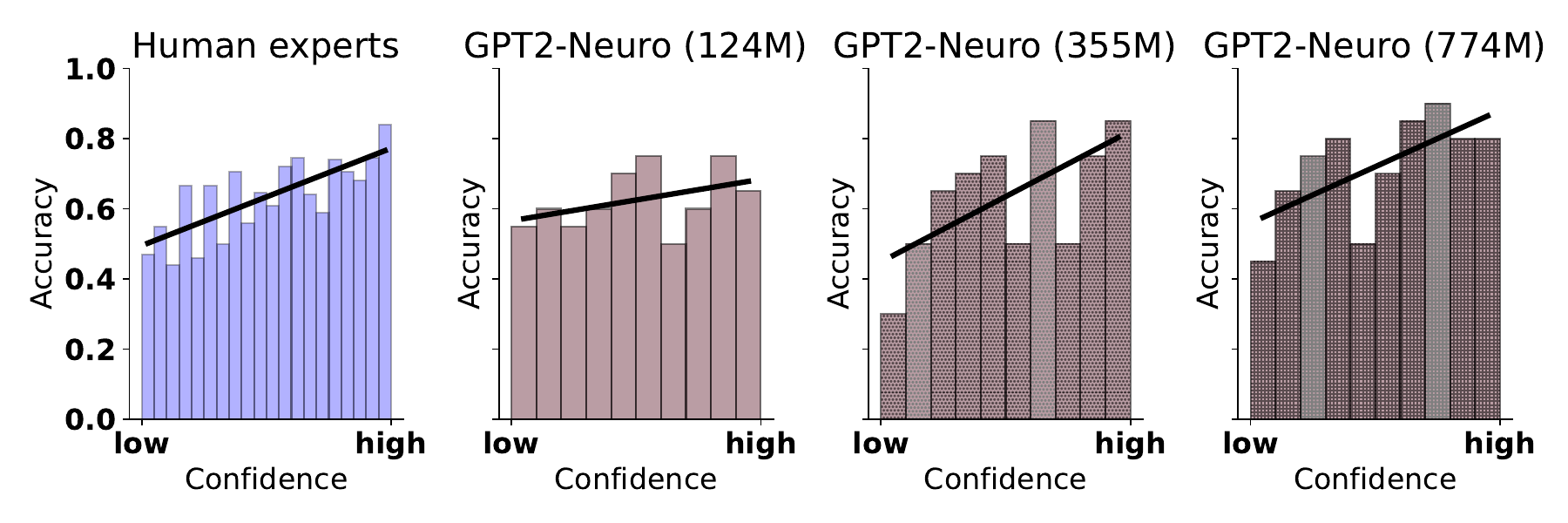}
    \caption{\textbf{Accuracy and confidence are calibrated for human experts and GPT-2 variants pretrained on neuroscience literature.} When human experts and models are confident in their BrainBench judgments, they are more likely to be correct. Confidence ratings were sorted and placed in equally-sized bins with the mean accuracy for items in that bin plotted. The positive slope of the black regression lines for human experts and all models indicates that confidence is well calibrated (i.e., higher confidence corresponds to higher accuracy). Calibration is beneficial for human-machine teams.}
    \label{fig:calibration}
\end{figure}

The absolute difference of perplexities of two versions of the abstract was used as a measure of model confidence. To assess the calibration of GPT-2 variants pretrained on neuroscience literature, we compared their accuracies with their confidence levels. First, we ranked and sorted model confidence across all test cases. Subsequently, we created 20 bins based on this sort. Within each bin, we calculated the mean accuracy. A well-calibrated model will exhibit a higher accuracy in bins associated with higher confidence rankings. We fit a linear regression model using the bin number as the independent variable and the mean accuracy of each bin as the dependent variable to evaluate calibration. We observed that human experts and models all show positive correlations between confidence and accuracy, indicating calibration (Fig. \ref{fig:calibration})

In addition, we fitted logistic regressions using perplexity differences to models' answers and from self-reported confidences of human experts to their responses. We confirmed a positive and significant correlation between model (355M and 774M) perplexities and their answers as well as human confidences and their responses (Table \ref{tab:lr_calibration}). 

\begin{table}[H]
\centering
\begin{tabular}{@{}llll@{}}
\toprule
Model/Human                           & Coefficient           & Intercept             & Test Statistics        \\ \midrule
GPT2-Neuro (124M)                     & $0.04 \pm 0.02$       & $0.41 \pm 0.05$       & $t(4)=2.14$, $p=0.049$ \\
GPT2-Neuro (355M)                     & $0.60 \pm 0.03$       & $0.23 \pm 0.05$       & $t(4)=19.95$, $p<.001$ \\
GPT2-Neuro (774M)                     & $0.87 \pm 0.09$       & $0.57 \pm 0.04$       & $t(4)=9.21$, $p<.001$ \\
Human experts                         & $0.01 \pm 0.00$       & $0.04 \pm 0.03$       & $t(4)=19.11$, $p<.001$ \\ \bottomrule
\end{tabular}
\caption{\textbf{Calibration analysis using logistic regression fits.} For models, logistic regressions were fitted between perplexity differences of a test case and its correctness given a LLM. For human experts, logistic regressions were fitted between their confidence of a test case and its correctness.}
\label{tab:lr_calibration}
\end{table}

\newpage
\section{Contextual integration analysis}
\label{app:context}
To investigate the extent to which GPT-2 models pretrained on neuroscience can integrate broad context from abstracts, we conducted an experiment involving the removal of contextual information from BrainBench test cases. Following the same evaluation procedure outlined for full abstract cases, we assessed the models using individual sentences extracted from abstracts containing at least one result alternation. In cases with multiple alternations, we computed the mean accuracy across these alternations as the final accuracy for the abstract. We then compared the level of performance degradation when these models were evaluated on full-length abstracts versus individual sentences where background and method information from the abstracts were removed.

\begin{figure}[H]
    \centering
    \includegraphics[scale=0.7]{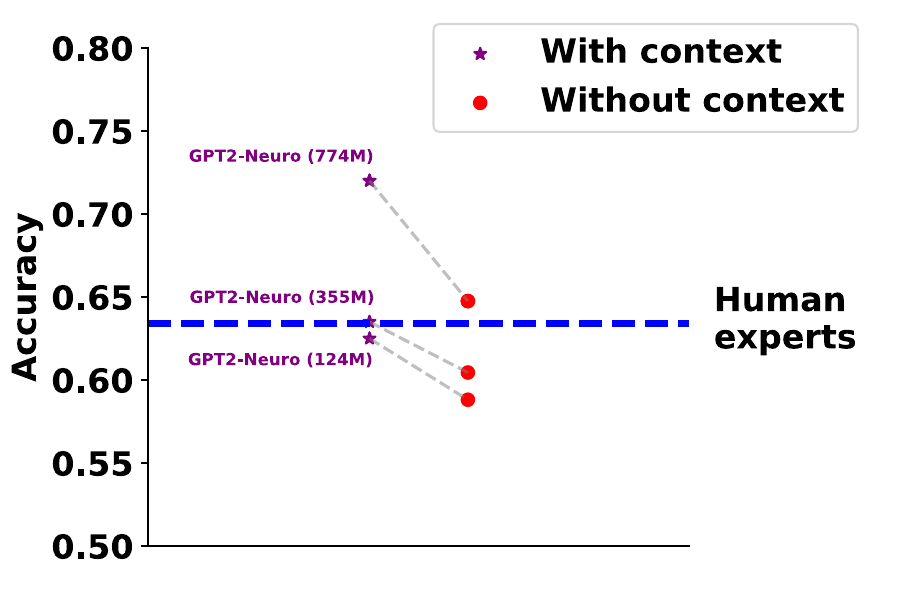}
    \caption{\textbf{GPT-2 variants pretrained on neuroscience literature integrate contextual information to succeed on BrainBench}. The removal of background and method sections from abstracts, with an evaluation based solely on individual sentences and result alternations, significantly impairs the performance of models on BrainBench. Much of the human-level performance appears to arise from integrating information across the abstract.
}
    \label{fig:iso}
\end{figure}

As a result, models performed worse when wider context was removed (Fig. \ref{fig:iso}), which provides strong evidence that these models are integrating information across the abstract, including information on background and methods.

\newpage
\section{BrainBench performance across model sizes and training data}
\label{app:overall}
\begin{figure}[H]
    \centering
    \includegraphics[scale=0.5]{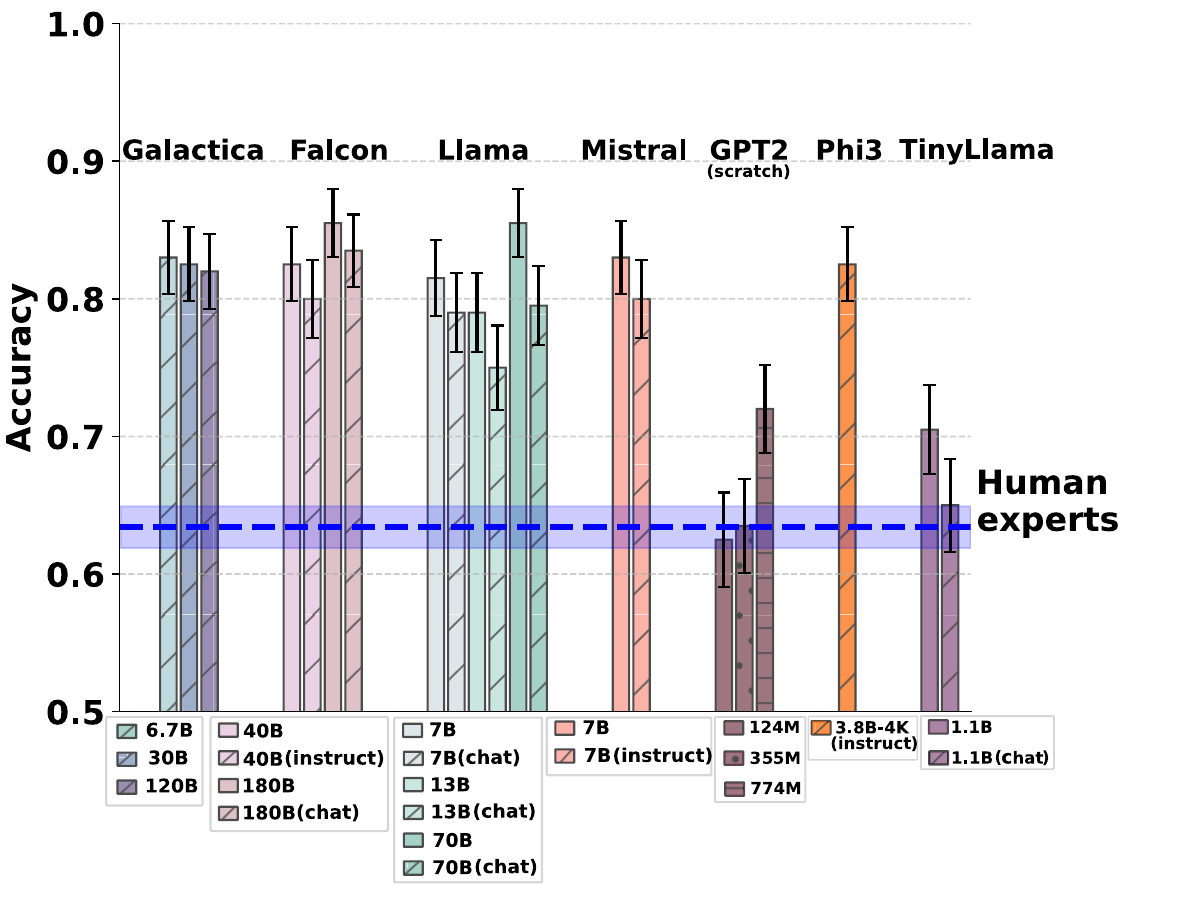}
    \caption{{\textbf{BrainBench performance across models of varying sizes and training data.} The GPT-2 variants (124M, 355M, 774M) pretrained entirely on the neuroscience literature from scratch shows progressively better results, matching or surpassing human performance and closing the gap to larger models tested in \citet{luo_large_2024}. Phi3, with half the size of the 7B models, achieves competitive results likely due to its high-quality training data. In contrast, TinyLlama, with 1.1 billion parameters, lags behind on BrainBench. Overall, performance on domain-specific tasks such as BrainBench is influenced by both model size and the quality and relevance of the training data.}}
    \label{fig:more}
\end{figure}

\section{Training details}
Variants of GPT-2 models using Huggingface implementations. We used a batch size of 16 for GPT-2 124M (8 for GPT-2 355M and 4 for GPT-2 774M) and a chunk size of 1024. Training involved the use of the AdamW optimizer \cite{loshchilov_decoupled_2019} with a learning rate of 2e-5 and a cosine learning rate scheduler. We applied gradient accumulation steps set at 8. Five training epochs were performed, along with a warm-up step of 0.03 and a weight decay rate of 0.001. bf16 mixed precision training and data parallelism were employed. We used 4 Nvidia A100 (80GB) GPUs hosted on Microsoft Azure.

\end{document}